\documentclass[english]{article}

\usepackage{geometry} 
\usepackage{graphicx}
\usepackage{amssymb}
\usepackage{babel}
\usepackage{epsfig}

\begin{document}

\begin{flushright} 
Bologna, \today \\ 
\end{flushright}

\vspace{0.8cm}

\begin{center}
{\Large {\bf OPAL and More}}

\vspace{0.6cm}

\normalsize {\bf The Benjamin C. Shen Memorial Symposium, UCR, March 8, 2008} 
\vspace{0.8cm}

\normalsize{
G. Giacomelli 

{\small
Physics Department of the University of Bologna and INFN Sezione di Bologna, I-40127 Bologna, Italy}
giacomelli@bo.infn.it
}
\vspace{0.5cm}

\normalsize{
P. Giacomelli 

{\small
INFN Sezione di Bologna, I-40127 Bologna, Italy}\\
paolo.giacomelli@bo.infn.it
}

\end{center}

\vspace{0.5cm}

{\bf Abstract.} {\normalsize 
We shall summarize some of the research activities performed in collaboration with Ben Shen in the OPAL experiment at LEP and in the CMS experiment at the LHC. And we shall recall the LEP legacy to particle physics in general and to the Standard Model in particular. Short recollections are made in other fields in which Ben was interested, in particular in Astroparticle Physics.}

\section{Introduction}

 One of the authors of this note (gg) met Ben Shen when he was in the USA (at BNL and then in Fermilab) in the 1960's; the acquaintance became a friendship, in particular when he was a visiting professor at the University of California at Riverside (UCR) in the winter quarter of 1970. Then in the early 1980's we started to collaborate in the OPAL experiment at LEP, where both were involved in the construction of the hadron calorimeter and both were members of the governing body of OPAL (at the beginning 6 people, later 9 people, and then all the group leaders of the collaborating Institutions). And both where among the initiators of the CMS collaboration at the LHC. The second author [pg] accepted a post doc position in the UCR group of Ben in the mid 90's to work in both the OPAL and CMS experiments at CERN; and later he was Adjunct Professor at UCR. 

 Both authors knew well the exceptional qualities of Ben, his excellent knowledge of physics, his leadership, his humanity and appreciated his friendship.

 In this note we shall recall some of the research work performed in collaboration in the construction and in the exploitation of the OPAL experiment at LEP, summarizing some of the main experimental results of OPAL and of LEP; some results obtained in direct collaboration are underlined. We shall then recall some of the activities performed in the construction of the CMS experiment at the LHC, summarizing some of the expected research fields. 

 Ben was very much interested in many topics of physics, as is well summarized in the contributions of the other speakers at this memorial symposium [http://www.physics.ucr.edu/Shen\_Symposium/shen\_ memorial.html] \cite{a}. Ben was always asking and discussing physics news in different fields, and proposed many seminars and colloquia at UCR and in other places, on the results of hadron-hadron collisions at the highest energy proton-synchrotrons, and later in astroparticle physics, which was a fast developing new field, in particular at the Gran Sasso Laboratories in Italy.

\section{The OPAL experiment at the LEP 
\boldmath{$e^{+}$$e^{-}$} collider}

 Fig. 1 shows a general schematics of the OPAL detector: like the other 4$\pi$ general purpose detectors at LEP, OPAL was made by several subdetectors, all with a cylindrical symmetry about the directions of the positron and electron colliding beams. The combined role of the subdetectors was to measure the energy, direction, charge and type of every produced particle. Each subdetector had a cylindrical structure with a ``barrel" and two ``end-caps". Tracking was performed by a central detector, including a microvertex silicon subdetector; electrons and photons were measured by electromagnetic calorimeters; the magnet iron yoke was instrumented as a hadron calorimeter and was followed by a muon detector \cite{1}. A forward detector completed the e.m. coverage and was used as a luminosity monitor: the luminosity of LEP was measured to a precision of about 0.3$\%$. The good quality of the subdetectors, the relatively low event rate and the very low background allowed to study in detail all the particles produced in each collision, with the exception of produced neutrinos (and neutralinos), which could only be inferred indirectly from missing mass and missing energy. 

The groups from the Universities of Bologna, of California at Riverside (UCR) and of Maryland worked together in the construction of the hadron calorimeter. The plastic limited streamer tubes were made in the INFN laboratories of Frascati, were brought to CERN for adding the strips (by the UCR group) and for testing and final assembling. We also contributed to other items, like the forward detector and the luminosity monitor. Fig. 2 shows a photo of early OPAL collaborators \cite{2} and of the main parts of the hadron calorimeter. 
        
 In its first phase of operation, LEP, called LEP1, functioned at c.m. energies around the Z$^{0}$ mass ($\sim$91 GeV). In the LEP2 phase the c.m. energy was gradually increased up to $\sim$210 GeV and this allowed to study the triple bosonic vertex Z$^{0}$W$^{+}$W$^{-}$ and performing searches for new particles of higher mass.

\begin{figure}[!ht]
\begin{center}
\vspace{1cm}
\includegraphics[width=.55\textwidth]{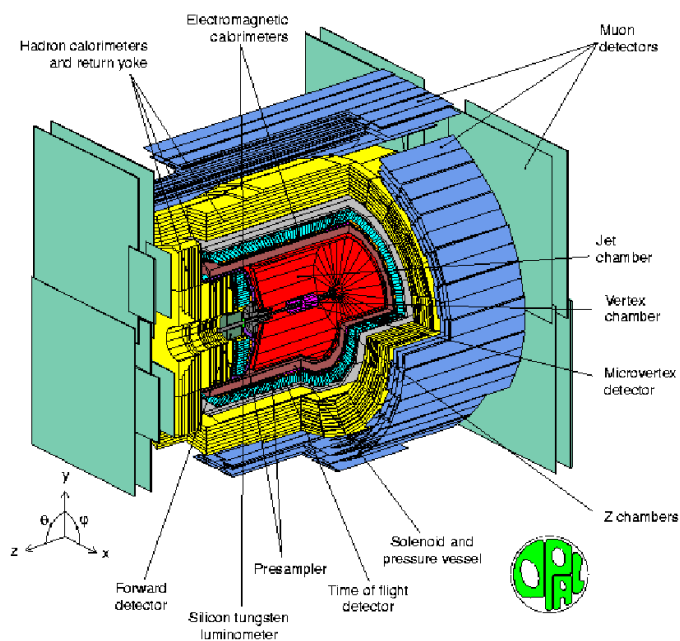}
 \begin{quote} 
\caption{\small The OPAL detector at LEP.}
\label {fig:1}
\vspace{-0.5cm}
\end{quote}
\end{center}
\end{figure}

\begin{figure}[!ht]
\begin{center}
\includegraphics[width=.50\textwidth]{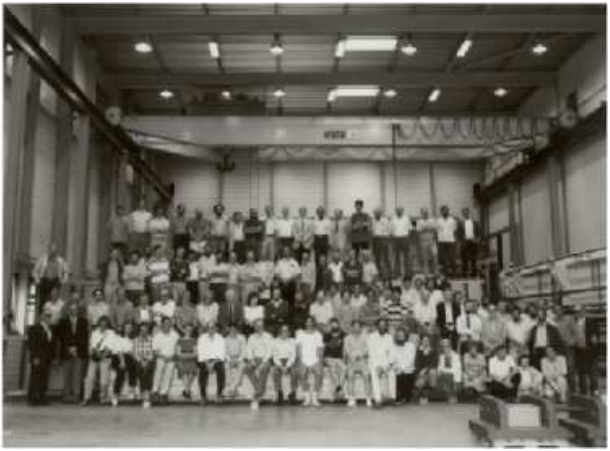}
\hspace{1cm}
\includegraphics[width=.42\textwidth]{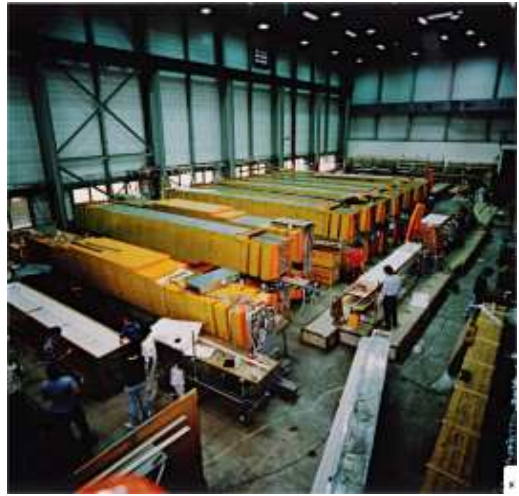}
\begin{quote} 
\caption{\small a) A photo of the OPAL collaborators. b) The OPAL hadron calorimeter in the assembly hall.}
\label {fig:2}
\vspace{-0.5cm}
\end{quote}
\end{center}
\end{figure}

 One of the main results of the experiments at LEP was to check the validity of the Standard Model (SM) of Particle Physics, of the Electroweak (EW) and Strong Interactions (SI), tested  to an unprecedented level of precision \cite{3}. We recall here that the basic components of the SM are quarks and leptons, which appear in three families: the first with the $u$, $d$ quarks and the $e^{-}$, $\nu_{e}$ leptons, the second with $c$, $s$ and $\mu^{-}$, $\nu_{\mu}$, the third with $t$, $b$ and $\tau^{-}$, $\nu_{\tau}$. The neutrinos were originally assumed to be massless and left handed, but the evidence for neutrino oscillations indicated that neutrinos have masses, although very small. Other fundamental objects of this SM are the force carriers, which are bosons of spin 1; they are the massless photon, carrier of the Electromagnetic Interaction (QED), the 8 massless gluons for the Strong Interaction (QCD), and the heavy carriers $Z^{0}$, $W^{+}$, $W^{-}$ for the Weak Interaction (WI). The number of constituents and carriers is large and one has also to consider the corresponding antiquarks and antileptons. The SM theory requires also a scalar boson, the $Higgs$ $boson$, needed for the spontaneous breaking of the Electroweak Symmetry and for the generation of masses.
   
{\bf Number of (neutrino) families.} At energies around the $Z^{0}$ peak the basic interaction processes are
\begin{eqnarray} 
e^{+} e^{-} \rightarrow Z^{0}, g \rightarrow f \overline{f}  \qquad     
  (f \overline{f} = q \overline{q}, l \overline{l})
\end{eqnarray}
and the behavior of the cross section is typical of a resonant state with J=1, described by a relativistic Breit-Wigner formula, which depends on the $Z^{0}$ mass m$_{Z}$ and on the width $\Gamma_{Z}$. The $Z^{0}$ decays ``democratically" into all particles and thus its width depends on the number of families. Fig. 3a and 3b show the combined data at LEP on the hadronic cross sections, e$^{+}$ e$^{-}$ $\rightarrow$ hadrons, from which one obtains the number of neutrino families, equal to 3 (more precisely N=2.9840$\pm$0.0082) \cite{3}.    

\begin{figure}[!ht]
\begin{center}
\vspace{-0.5cm}
\includegraphics[width=.53\textwidth]{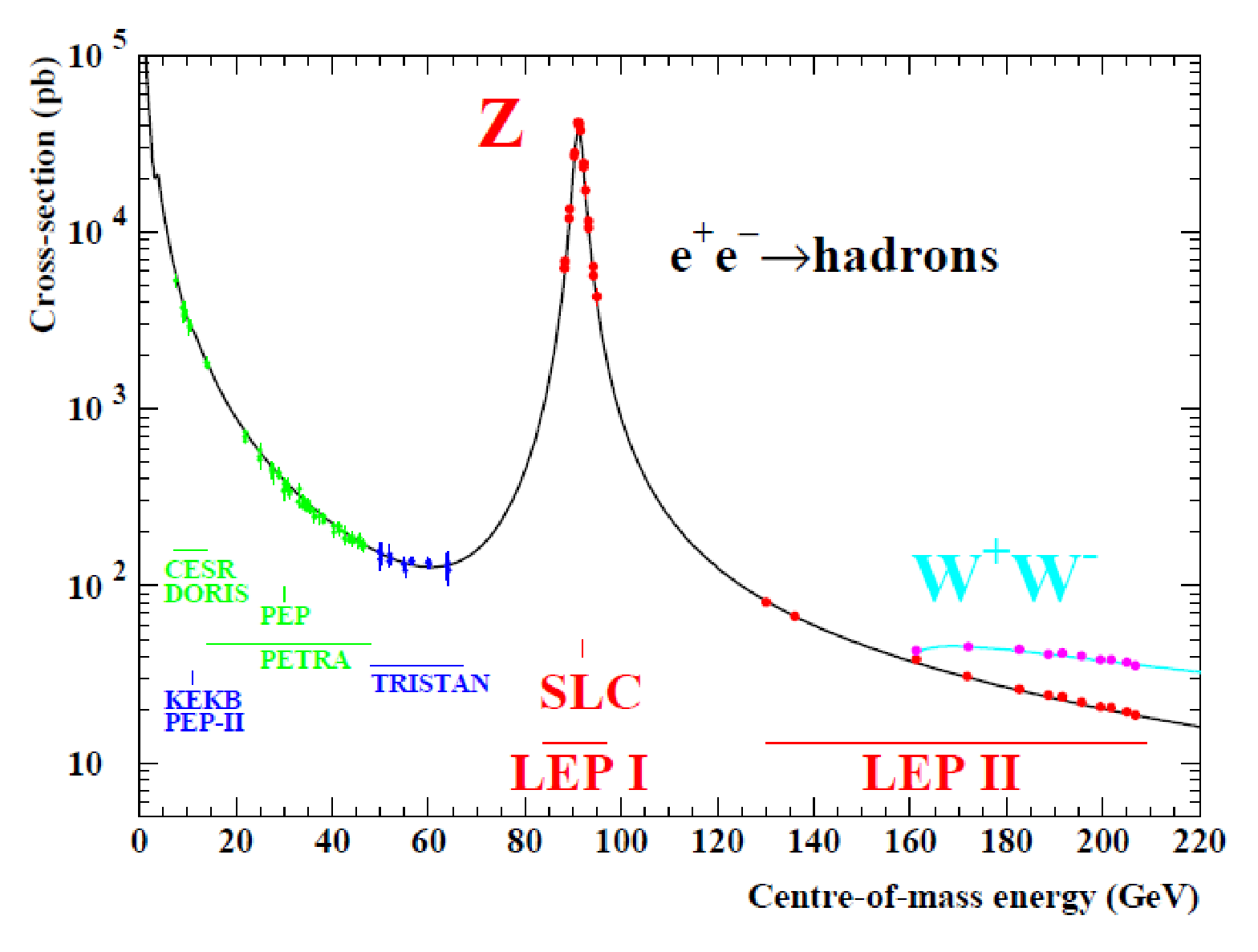}
\hspace{1cm}
\includegraphics[width=.38\textwidth]{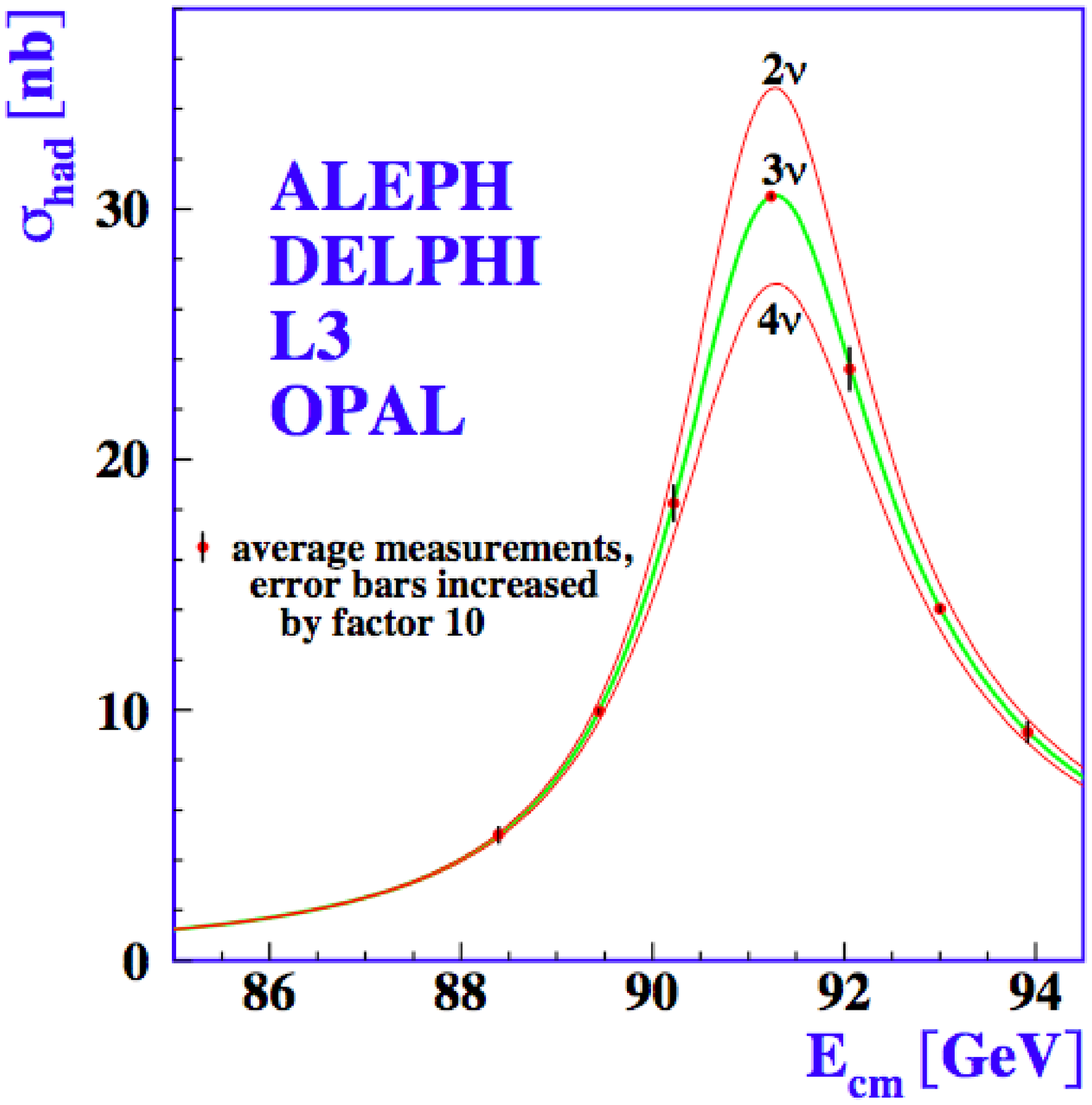}
\begin{quote} 
\caption{\small a) The hadronic $e^{+}$$e^{-}$ cross-section as a function of centre-of-mass energy The solid line is the prediction of the SM, and the points are the experimental measurements. Also indicated are the c.m. energy ranges of various $e^{+}$$e^{-}$ accelerators. The cross-sections have been corrected for the effects of photon radiation. b) Number of neutrino families.}
\label {fig:3a3b}
\vspace{-0.5cm}
\end{quote}
\end{center}
\end{figure}

\begin{figure}[!ht]
\begin{center}
\vspace{-0.5cm}
\includegraphics[width=.54\textwidth]{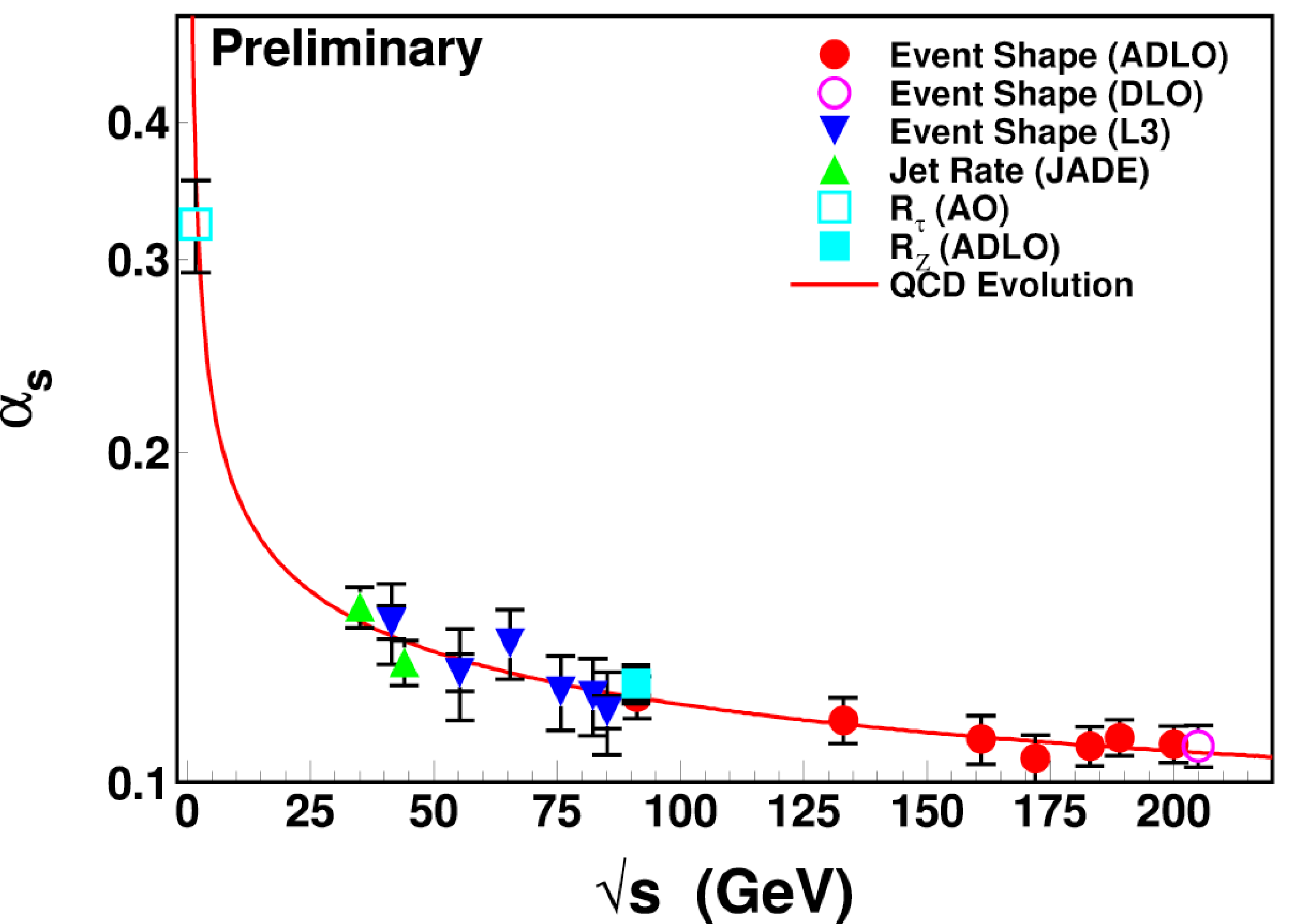}
\includegraphics[width=.44\textwidth]{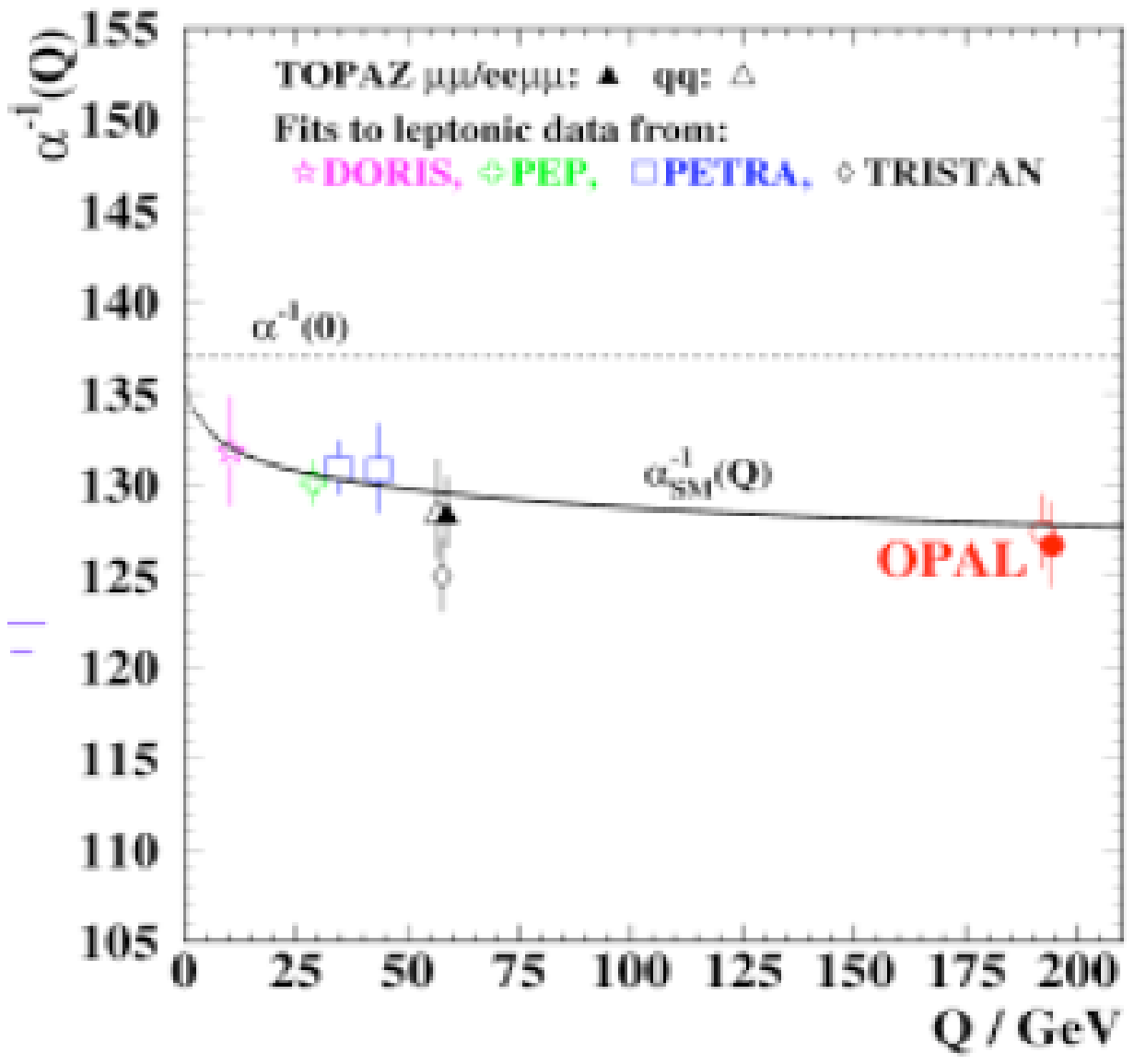}
\begin{quote} 
\caption{\small a) Decrease of the strong coupling constant with increasing energy. b) Also $\alpha_{EM}$ varies with energy.}
\label {fig:4a4b}
\vspace{-0.5cm}
\end{quote}
\end{center}
\end{figure}

\begin{figure}[!ht]
\begin{center}
\vspace{-0.5cm}
\includegraphics[width=.60\textwidth]{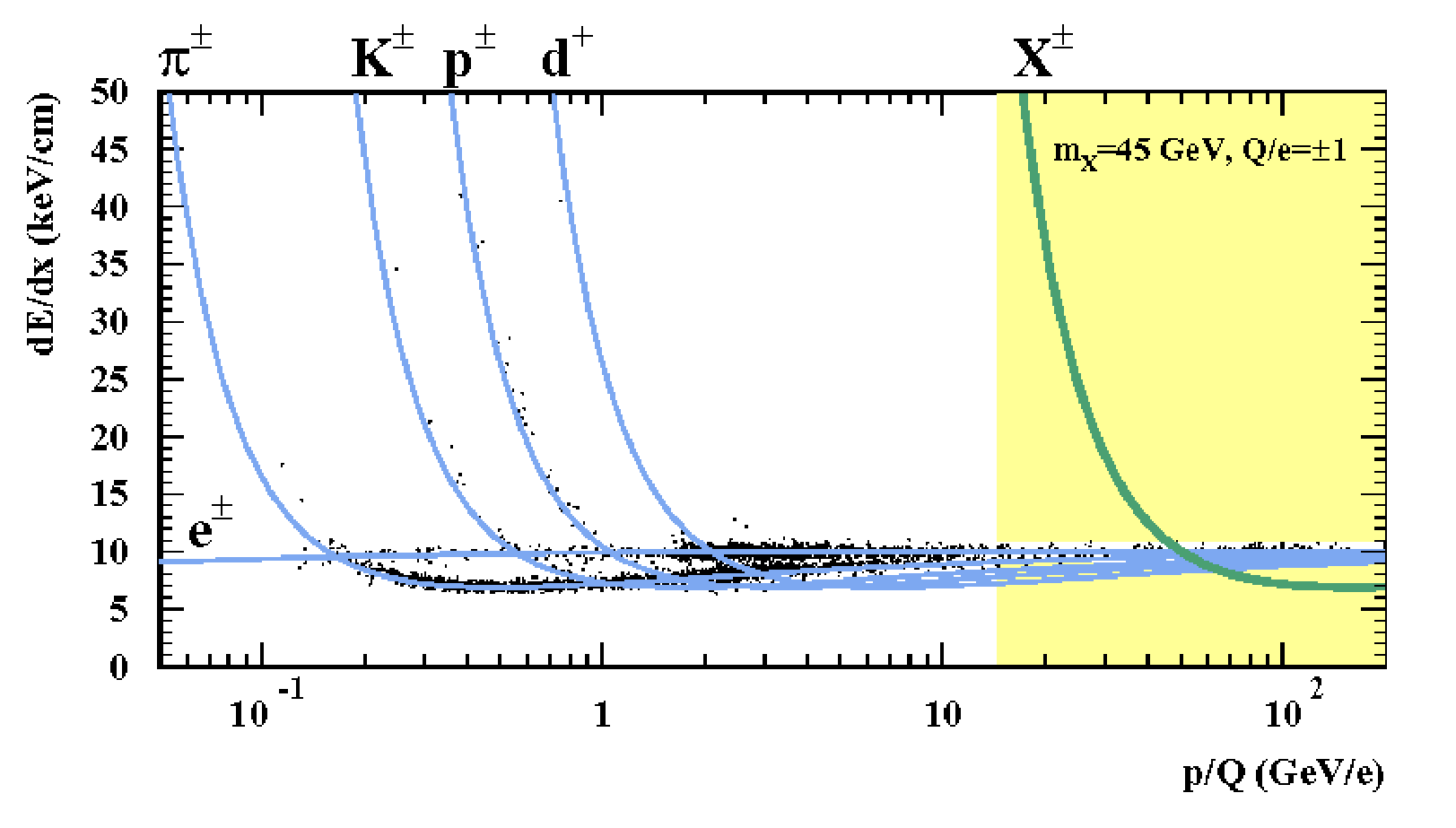}
\includegraphics[width=.35\textwidth]{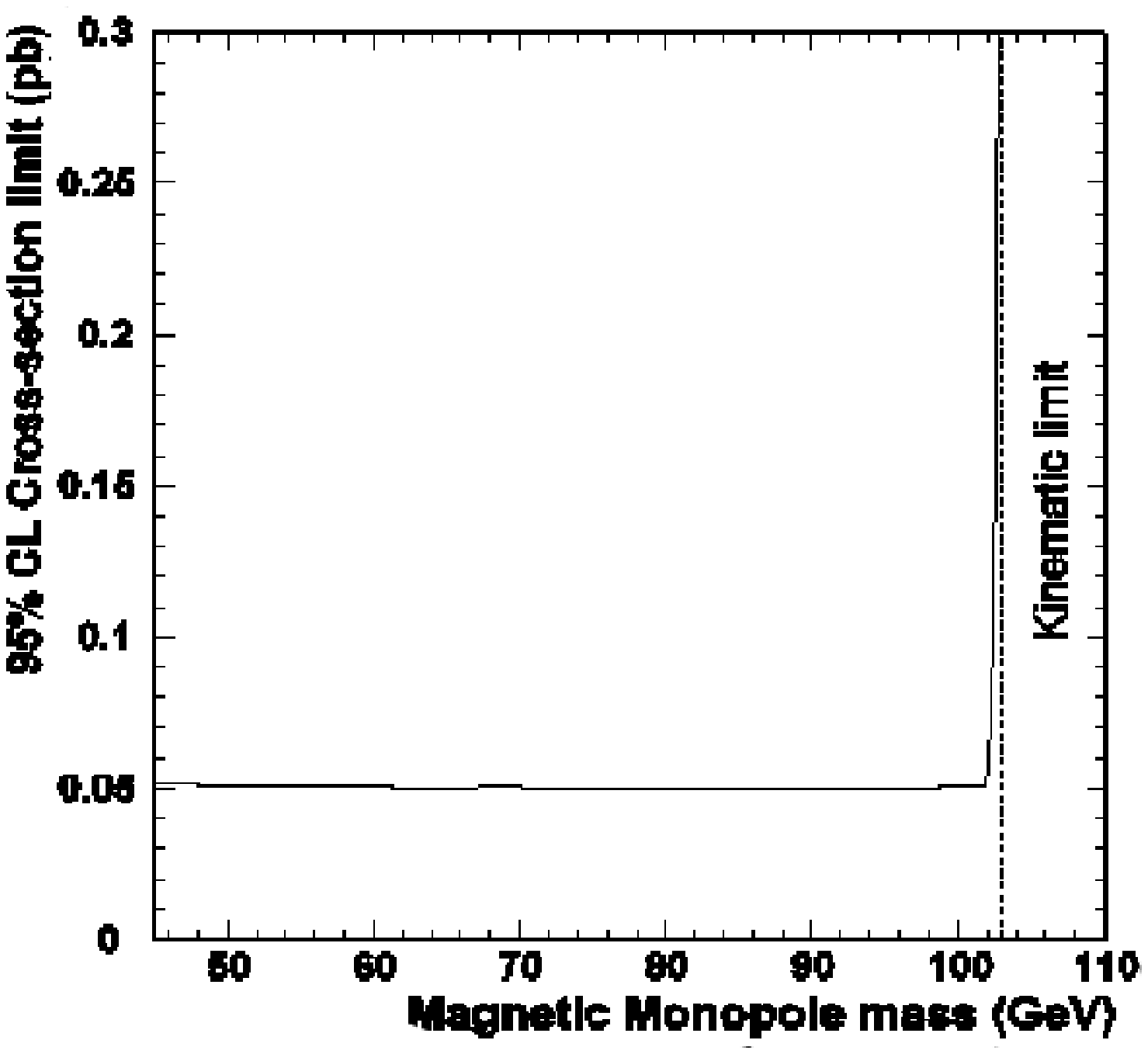}
\begin{quote} 
\caption{\small a) Searches for stable heavy charged particles. b) Upper limit for the production cross section of Dirac magnetic monopoles.}
\label {fig:5a5b}
\vspace{-0.5cm}
\end{quote}
\end{center}
\end{figure}

\begin{figure}[!ht]
\begin{center}
\vspace{-0.5cm}
\includegraphics[width=.52\textwidth]{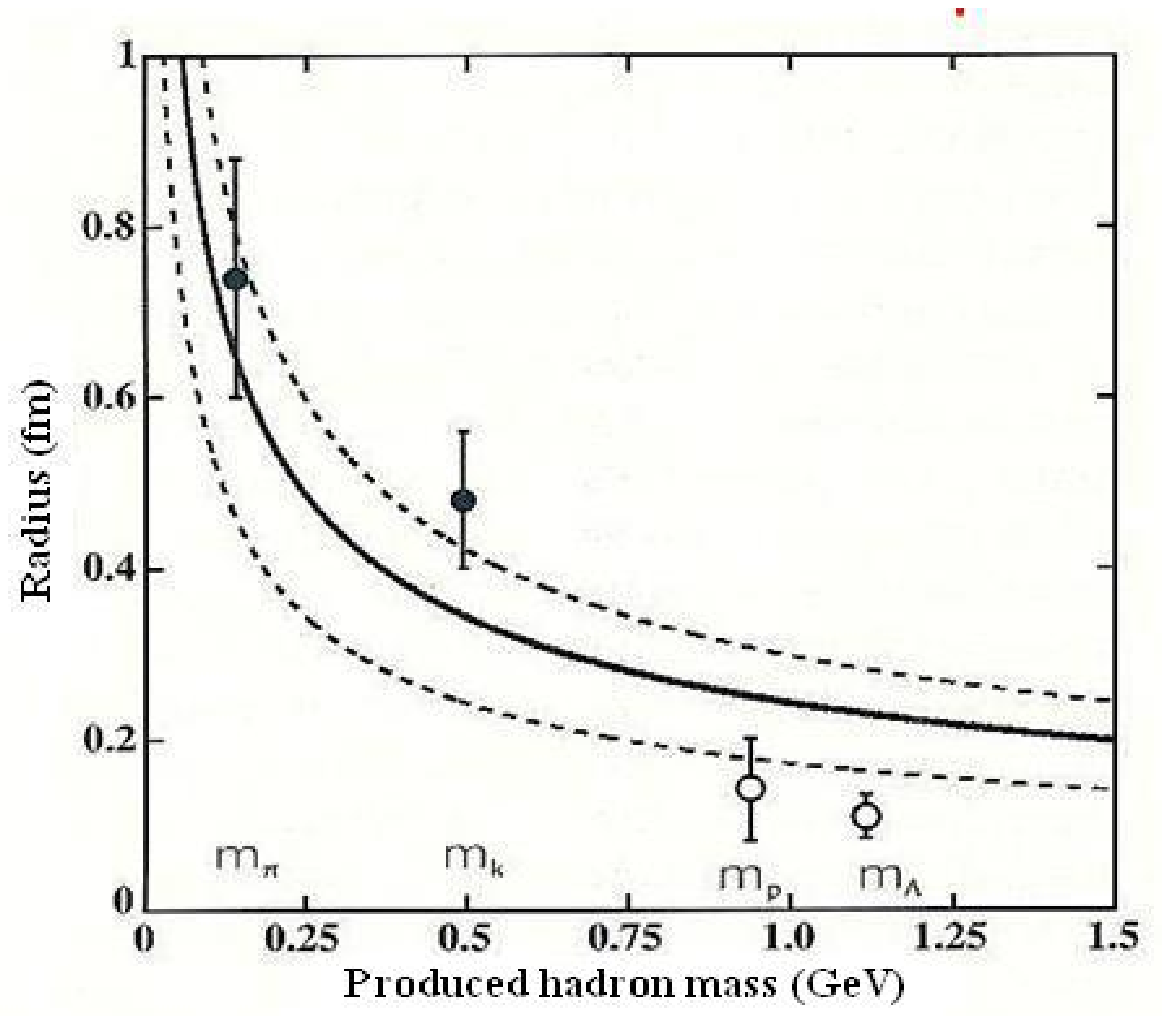}
\hspace{-0.3cm}
\includegraphics[width=.42\textwidth]{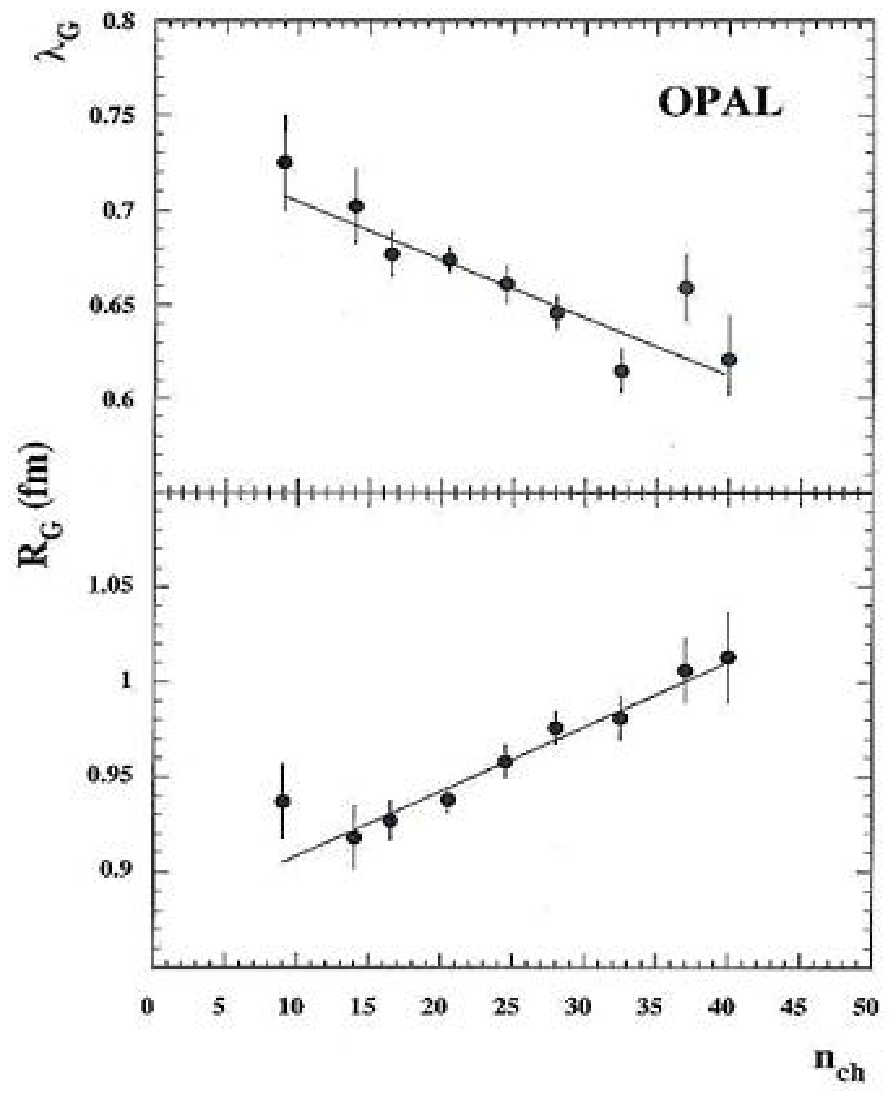}
\begin{quote} 
\caption{\small Bose-Einstein correlations: a) Static radius vs mass of produced particles. b) Dependence of the static radius on the charged event multiplicity.}
\label {fig:6a6b}
\vspace{-0.5cm}
\end{quote}
\end{center}
\end{figure}

\begin{figure}[!ht]
\begin{center}
\includegraphics[width=.47\textwidth]{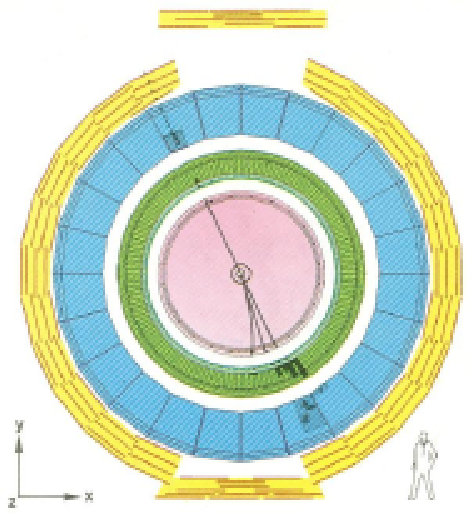}
\hspace{-1cm}
\includegraphics[width=.57\textwidth]{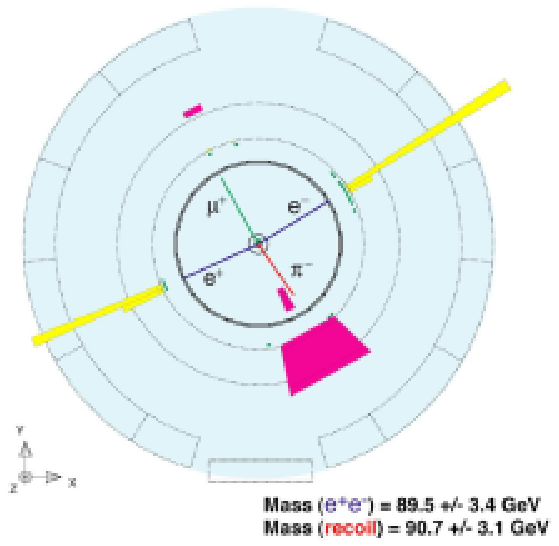}
\begin{quote} 
\caption{\small a) One OPAL event at LEP1, $e^{+}$$e^{-}$ $\rightarrow$ $q\overline{q}$ $\rightarrow$ 2 jets of hadrons. b) One event at LEP2, $e^{+}$$e^{-}$ $\rightarrow$ $Z^{0}$$Z^{0}$ $\rightarrow$ $e^{+}$$e^{-}$ $\tau^{+}$$\tau^{-}$ ($\tau^{+}$$\rightarrow$$\mu^{+}$$\nu_{\mu}$$\overline{\nu_{\tau}}$,  $\tau^{-}$$\rightarrow$$\pi^{-}$$X$).}
\label {fig:7}
\vspace{-0.8cm}
\end{quote}
\end{center}
\end{figure}

{\bf The strong coupling constant.} The measurement of the ratio of the number of three hadron jets to the number of two hadron jets at any energy is one way to measure $\alpha_{S}$, $the$ $strong$ $coupling$ $constant$. This is a fundamental parameter of QCD. It was measured in many precise different ways and it was found that \cite{4}:

 i) $\alpha_{S}$ is flavour independent,

 ii) it decreases with increasing energy ($running$ $of$ $the$ $strong$ $coupling$ $constant$), see Fig. 4a. 

 Also the $electromagnetic$ $coupling$ $constant$, $\alpha_{EM}$, is not constant: it increases from the value $\alpha_{EM}$ (at rest)$\sim1/137$ at zero energy to $\alpha_{EM}$(m$_{Z}$)$\sim1/128$, see Fig. 4b \cite{5}. 

 A large variety of phenomenological studies were made on QCD, including the first determination below threshold of the $b$ quark mass \cite{2}, the complexity of the hadron spectrum, phase transitions, the shape of quark and gluon jets, etc \cite{6}. In the past the reference theory was considered to be QED, but now a preference is expressed for QCD.

 {\bf Searches for heavy stable charged particles}. One of the searches performed in common between Bologna and UCR was the search for new heavy stable particles. The calibration of the method was performed with the production of K$^\pm$, protons and antiprotons, deuterons and antideuterons and was based on the very good performances of the central detector, in particular of its exceptional energy loss resolution. Good limits were established up to the energy limit from the available LEP energies, Fig. 5a \cite{6}. The method was extended to the search for pair produced Dirac Magnetic Monopoles up to Magnetic Monopole masses of $\sim$100 GeV, Fig. 5b \cite{7}.

 {\bf Bose-Einstein Correlations (BECs)} are a quantum mechanical phenomenon which manifests in final multihadron states as an enhanced probability for identical bosons to be emitted with small relative four momentum Q compared with non identical bosons under similar kinematic conditions. From the measured effect it is possible to determine the space-time dimensions of the boson-emitting source. The effect arises from the ambiguity of path between sources and detectors and the requirement to symmetrize the wave function of two or more identical bosons. The analyses involved first a static source for which we determined its emitting radius ($\sim$1 fm for pions, $\sim$0.7 fm for kaons), the increase of the emitting radius with increasing event multiplicity and a smaller radius for three pion correlations with respect to two pion correlations, Fig. 6. BECs in two and three dimensions allowed to observe that the longitudinal radius is about 20$\%$ larger than the transverse radius, indicating that the source is elongated in the q$\overline{q}$ direction. The BECs were then analyzed in a non static situation using specific models and we found a more complex view indicating an expanding source, like for ion-ion collisions. Expanding sources may arise in positron-electron collisions because of string fragmentation. This example of analysis reveals the complexity and the richness of multihadron analyses \cite{8}. 

 Many studies involved the analysis of gluon jets \cite{10n}.

 Fig. 7 shows some special events in the OPAL detector at LEP1 and at LEP2 energies: they were often used in formal lectures and for outreach purposes.

\section{The CMS experiment at the LHC}

 In the 1990's most US high energy physicists started to be involved in the Superconducting Super Collider (SSC) in the US. Ben Shen preferred to start preparatory work for a possible experiment at the CERN Large Hadron Collider (LHC), which was slowly going ahead. When the SSC was canceled in 1994 and then the LHC was adopted by everybody, prof Shen was one of the founders of the Compact Muon Solenoid (CMS) experiment  at the LHC and played a major role in the US participation in CMS, and more generally in the LHC, which is located in the old underground LEP tunnel. The LCH is a superconducting $pp$ collider at a center of mass energy of 14 TeV; it should also accelerate heavy ions so as to have the possibility to study the transition to the quark gluon plasma at very high energies, in a large range of rapidity \cite{9}. 

 Ben and his group started working in the End Cap Muon Detector of CMS, while our Bologna group started working in the Barrel Muon Detector. 

 The CMS collaboration has grown to more than 2000 physicists and will run for the next decade. LHC experiments should start running at the end of 2009. CMS is one of the major general purpose experiments and has the general barrel-end cup structure of all experiments at high energy colliders; it is characterized by a very high field superconducting solenoidal magnet, with a field of 4 T at the center, and a stored magnetic energy of 2.6 GJ, see Fig. 8. Note also the very large structure of the experiment and that each one of its subdetectors has a very large number of electronic channels \cite{10}. Each $pp$ collision at the LHC will have a high multiplicity, 70-90 produced charged particles per $pp$ interactions, and several interactions during the time of each beam-beam encounter) \cite{11}.

 CMS will have several refined triggers and the analysis of the events will be very complex and will need the availability of an extremely large computing power: for this purpose the GRID system was invented and deployed: the most powerful unit, the so called GRID ``tier zero'', is at CERN, while in each participating nation is available one ``tier 1'' and smaller entities. At the moment they are all successfully used for Monte Carlo simulations and are getting ready to analyze real data.
 
 One of the first searches by all experiments at the LHC will be the search for the Higgs boson in this new energy region, and CMS is well placed for it. The search will include the SM Higgs and also the Higgs present in some Supersymmetric Models. The searches for physics beyond the SM, in particular Supersymmetry, will be part of the main topical searches, but also more standard topics will be considered. The collaboration is reviewing the performances of several types of triggers: the Higgs triggers, the heavy flavour triggers, and many others.   

\begin{figure}[!t]
\begin{center}
\includegraphics[width=.50\textwidth]{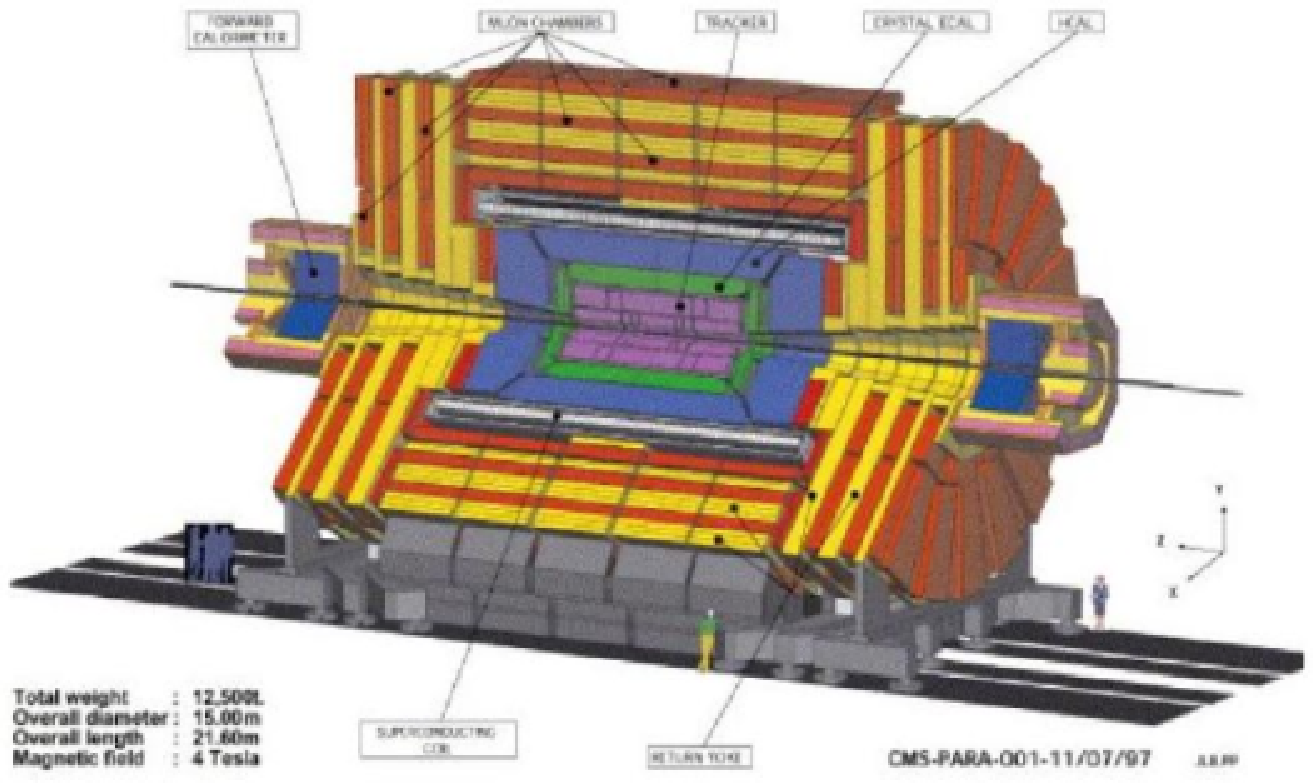}
\hspace{-0.5cm}
\includegraphics[angle=90, width=.49\textwidth]{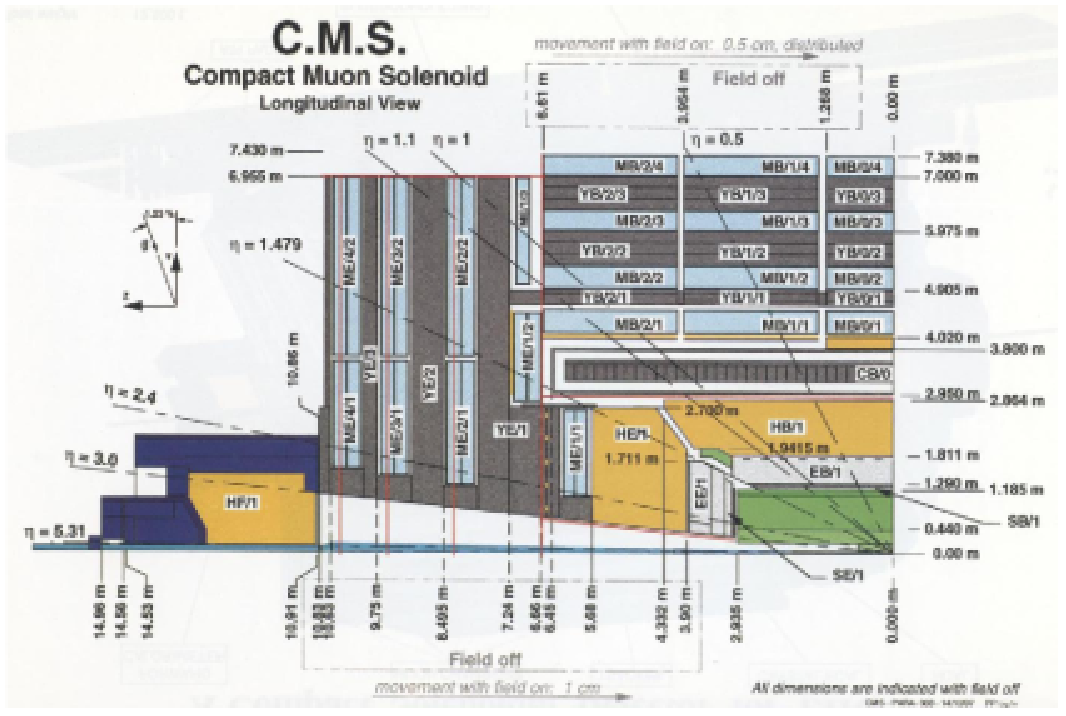}
\hspace{1cm}
\begin{quote} 
\caption{\small The CMS detector at the LHC. a) Global view, b) longitudinal view.}
\label {fig:8a8b}
\vspace{-0.5cm}
\end{quote}
\end{center}
\end{figure}

\section{Hadron-hadron collisions: Elastic scattering, Hadron production, Total hh cross sections, Antinuclei}

 These were the fields mostly studied at BNL, CERN, Fermilab, Serpukhov and other lower energy proton accelerator laboratories in the 1960s-1990s and will be studied again at LHC. 

 Hadron-hadron {\bf elastic scattering} involved first the study of low energy pion-nucleon scattering and of resonances \cite{12}; later it involved the study of the forward diffraction region of higher energy elastic scattering for the six stable charged hadrons ($\pi^{\pm}$, $K^{\pm}$, $p$ and $\overline{p}$ on protons \cite{13}): their interpretation was made in the context of Regge poles and trajectories, in particular of the Pomeron trajectory \cite{14}.  

 The {\bf study of the secondary beams} produced in the collision with several types of targets of the protons accelerated at proton synchrotrons of increasing energy was a prerequisite study before performing any other experiment; the beams were momentum selected by a number of magnetic dipoles, focused by quadrupoles, and defined by collimators and scintillation counters; the wanted particles in the beams were defined by scintillation counters and electronically selected by precise differential Cherenkov counters. These experimental data yielded interesting results on simple scaling laws for particle production \cite{15}.

 {\bf Total cross section measurements} were made in ``good geometry" using  the available secondary beams, which were well defined with scintillation counters and were well selected with differential gas Cherenkov counters;  standard targets of liquid hydrogen, liquid deuterium and several nuclear targets were available; the unscattered beam was counted by a series of circular scintillation counters of increasing radius. The total cross sections had statistical errors smaller than 0.3$\%$, while the systematic uncertainty was about 1-2$\%$. In the resonance region, for c.m. energies smaller than 3 GeV, the cross sections were characterized by the presence of peaks and structures \cite{16}; at higher energies most cross sections decreased regularly with increasing energy. At even higher energies followed the ``flattening'' of the energy dependence of negative pion-proton and negative kaon-proton total cross sections and then there was the unexpected discovery of the rising of the K+p total cross sections at Serpukhov \cite{19n}; later the increasing with energy of the pp total cross section was established at the CERN ISR and tentatively in cosmic rays \cite{20n}; and finally the rising of all hadron-hadron total cross sections \cite{21n} and of the p$\overline{p}$ cross sections were established at Fermilab \cite{22n}. It is still not completely clear why all cross sections increase with increasing energies, though it was suggested that it may be connected with the increasing cross section of mini-jets due to collisions with gluons inside the proton. These studies will continue exploiting the higher c.m. energies which will become available at the LHC.
 
 {\bf Production of antinuclei in a high intensity RF separated beam.} At the CERN SPS, in the 1970s, was constructed a special high intensity radiofrequency (RF) separated beam to study particle production in the very forward direction, to search for new massive particles and to study antinuclei production. The beam intensity was more than 2$\cdot$10$^{7}$ protons per pulse; the rejection of unwanted particles was achieved  first with the RF separators, and then electronically with several differential Cherenkov counters and with charge measurements in scintillators. Many thousand antideuterons and several hundred antitritons ($\overline{t}$) and antihelium-3 ($\overline{He}$) nuclei were recorded, see Fig. 9 \cite{23n}.  

\begin{figure*}[t]
\vspace{-1.5cm}
\begin{minipage}{0.5\linewidth}
\includegraphics[width=.83\textwidth]{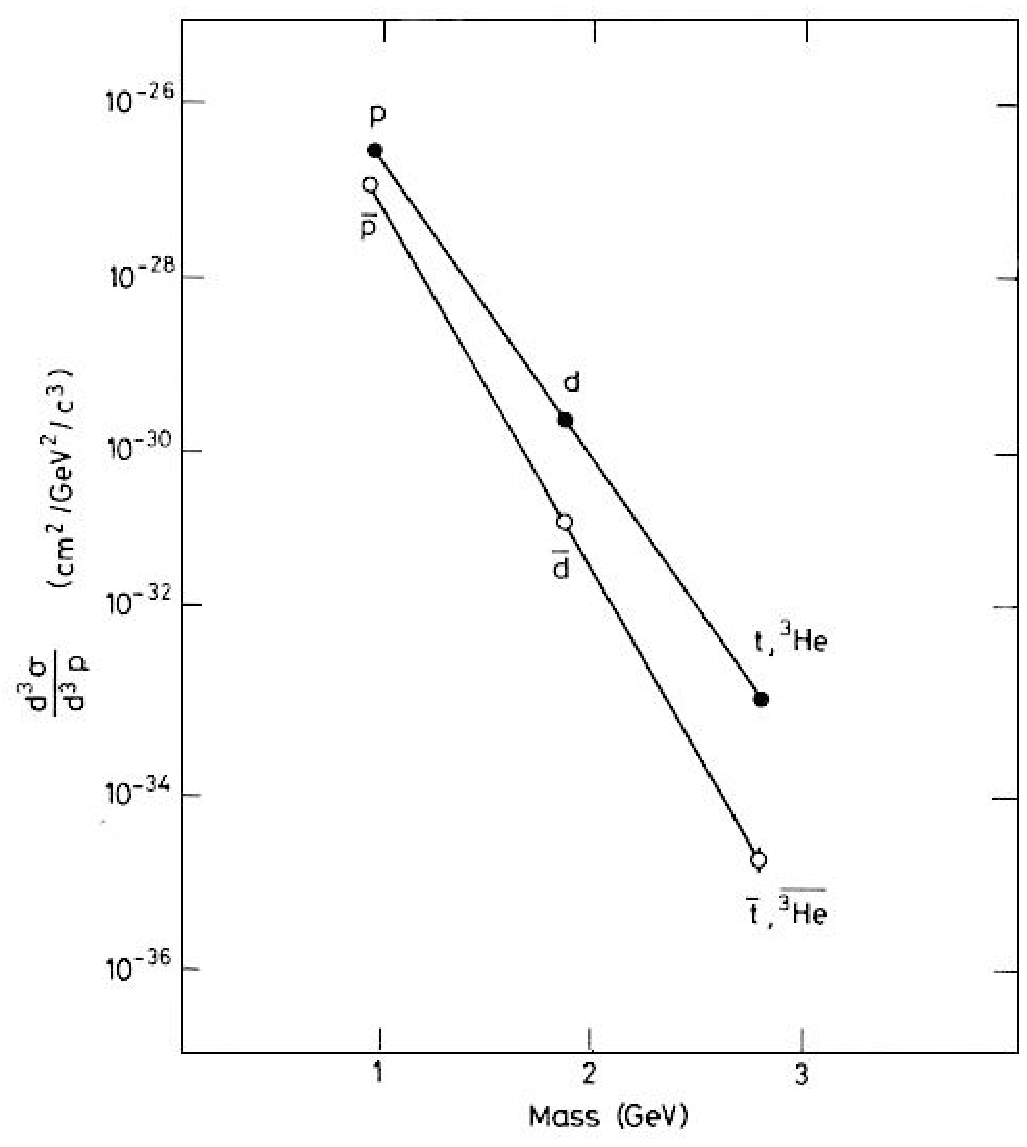}
\caption{\small CERN-SPS. Invariant cross sections for the productions of $p$, $d$, $t$, and $\overline{p}$, $\overline{d}$, $\overline{t}$, $\overline{He^{3}}$ at $x$$=$0 vs mass. The lines are fits to exponentials. }
\label{fig:zenith}
\end{minipage}\hfill
\begin{minipage}{0.45\linewidth}
\includegraphics[width=1\textwidth]{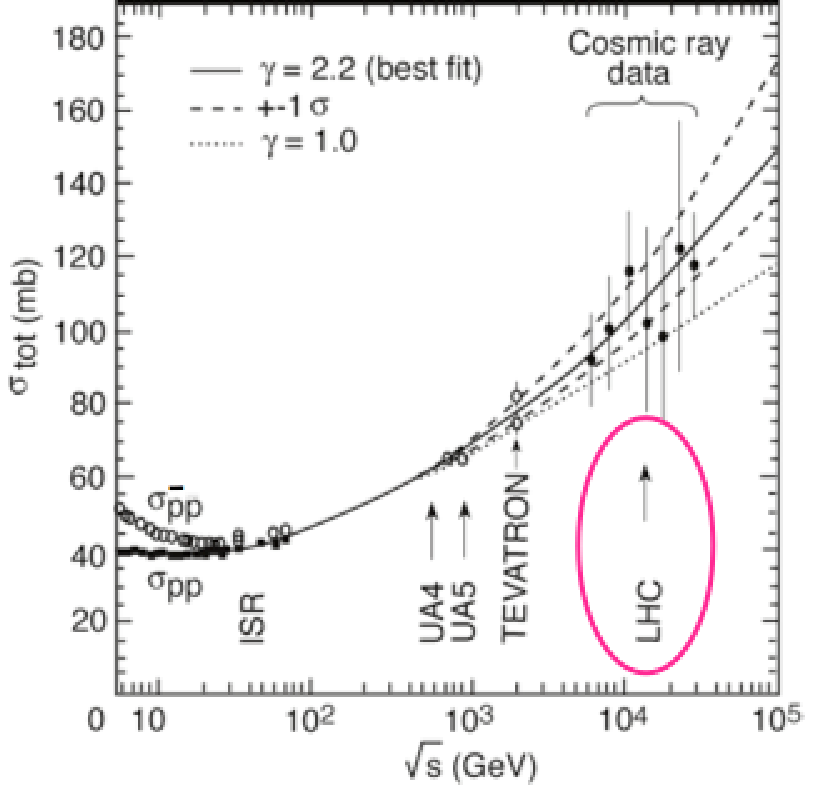}
\caption{\small Compilation of $\overline{p}p$ and $pp$ total cross sections vs c.m. energy; they include cosmic ray measurements.} 
\label{fig:le}
\end{minipage}
\end{figure*}

\section{Astroparticle physics}

 As already stated, Ben was interested in several new fields of physics, in particular  astroparticle physics. His interest ranged from the novelty of the field, to its connections with the particle field and with astrophysics and cosmology. He participated in the Milagro experiment \cite{a} and was involved in the search of a location for an underground lab in the US, following the example of the Gran Sasso lab in Italy, where experiments are performed on: solar and atmospheric neutrinos \cite{24n, 25o}, supernova neutrinos, nuclear astrophysics, high energy cosmic rays, dark matter, neutrinoless double beta decay and magnetic monopole searches \cite{25n}, and on the CERN to Gran Sasso neutrino beam (CNGS) \cite{26n}, Fig. 11a.
 
{\bf Neutrino masses.} Astrophysics informations yield an upper limit on the sum of the masses of the 3 neutrino types: $\sum$m$_{\nu}$$<$0.6 eV.

{\bf Neutrino oscillations.} If neutrinos have non zero masses, we have to consider the $weak$ $flavour$ $eigenstates$ $\nu_{e}$, $\nu_{\mu}$, $\nu_{\tau}$ and the $mass$ $eigenstates$ $\nu_{1}$, $\nu_{2}$, $\nu_{3}$. Neutrino flavour eigenstates are relevant in decays ($\pi^{+}$$\rightarrow$ $\mu^{+}$+$\nu_{\mu}$ and in interactions $\nu_{\mu}$+$n$ $\rightarrow$ $\mu^{-}$+$p$), while neutrino mass eigenstates are relevant in neutrino propagation. Flavour eigenstates may be written as linear combinations of mass eigenstates. For 2 flavours ($\nu_{\mu}$, $\nu_{\tau}$) and 2 mass eigenstates ($\nu_{2}$, $\nu_{3}$) we have
\begin{eqnarray}
\nu_{\mu}=\nu_{2}\,cos\Theta_{23}+\nu_{3}\,sin\Theta_{23}\nonumber \\
\nu_{\tau}=-\nu_{2}\,sin\Theta_{23}+\nu_{3}\,cos\Theta_{23}
\end{eqnarray}
where $\Theta_{23}$ is a mixing angle. The survival probability of a $\nu_{\mu}$ beam is
\begin{eqnarray}
P(\nu_{\mu}\to\nu_{\mu})=1-P(\nu_{\mu}\to\nu_{\tau})
\end{eqnarray}
The probability of $\nu_{\mu}$ $\rightarrow$ $\nu_{\tau}$ transition is (
$\Delta m^{2}_{23}$=m$^{2}_3$-m$^{2}_{2}$):
\begin{eqnarray}
P(\nu_{\mu}\to\nu_{\tau})= sin^{2}2\Theta_{23}\,sin^{2}\left(\frac{1.27\Delta m^{2}_{23}L}{E_{\nu}}\right)
\end{eqnarray}

{\bf Atmospheric neutrinos} are produced in the chain of interactions and decays of high energy cosmic rays (CRs) in the earth atmosphere: primary CRs, protons and nuclei, interact with air nuclei of the upper atmosphere producing pions and kaons, which decay into muons and $\nu_{\mu}$'s; then the muons decay yielding $\nu_{\mu}$ and $\nu_{e}$ (the final ratio of the numbers of $\nu_{\mu}$ to the number of $\nu_{e}$ is 2). Atmospheric neutrinos are effectively generated at 10-20 km above ground and proceed towards the earth. They have  energies ranging from a fraction of GeV to more than 100 GeV and travel distances from tens of km (downgoing neutrinos) to 13000 km (upgoing neutrinos). Atmospheric neutrinos are well suited to study neutrino oscillations for 10$^{-3}$$<$$\Delta m^{2}_{23}$$<$10$^{-2}$ eV$^{2}$. The experiments IMB and Kamiokande reported anomalies in the ratio of muon to electron neutrinos while MACRO reported a deficit of upthroughgoing muons \cite{24n}. In the neutrino '98 conference the experiments Soudan2, MACRO \cite{24n} and SuperK reported deficits in the atmospheric $\nu_{\mu}$ fluxes and angular distribution distortions with respect to non oscillated predictions. These features are explained in terms of $\nu_{\mu}$$\rightarrow$$\nu_{\tau}$ oscillations with $\Delta m^2_{23} \simeq 0.0024$ eV$^2$ and maximal
mixing.

{\bf Long baseline neutrino experiment} results gave further insight into neutrino physics. The first long baseline experiment was the K2K experiment (in the KEK to Kamioka beam in Japan), the second was the MINOS experiment (in the NUMI neutrino beam from Fermilab to the Soudan mine in the USA) \cite{25o}, the third experiment is the OPERA experiment (in the CERN to Gran Sasso CNGS neutrino beam); the first two experiments are $\nu_{\mu}$ disappearance experiments, while the third searches for $\nu_{\mu}$$\rightarrow$$\nu_{\tau}$ ($\nu_{\tau}$ appearance experiment). 
 OPERA at Gran Sasso is a hybrid emulsion-electronic detector; the $\nu_{\tau}$ appearance will be made by direct detection of the $\tau$ lepton from $\nu_{\tau}$ CC interactions and the $\tau$ lepton decay products. To observe the decays, a spatial resolution of $\sim$1 $\mu$m is needed; this is obtained in thin emulsion sheets interspersed with thin lead plates ``Emulsion Cloud Chamber" (ECC), assembled in ``bricks"); the electronic detectors are needed to find the brick in which the neutrino interaction occurred and to measure the muon momentum. A fast automated emulsion scanning system with a scanning speed of $\sim$20 cm$^{2}$/h is needed to cope with daily analyses of many emulsions. This is a factor of 10 larger speed compared to past systems. The main purpose of the Opera experiment is to prove that the oscillation phenomenon really exists in the form sketched above.
 
 \begin{figure}[!t]
\begin{center}
\includegraphics[width=.50\textwidth]{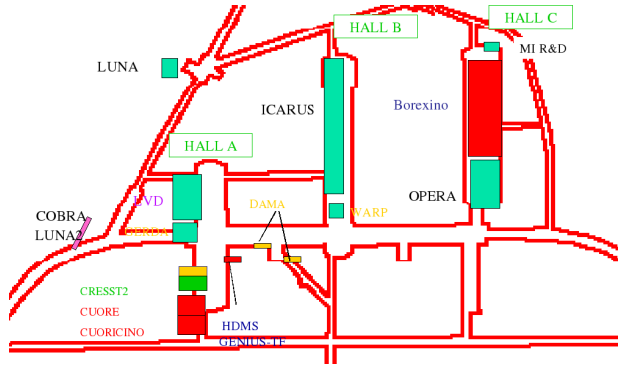}
\hspace{0.5cm}
\includegraphics[width=.45\textwidth]{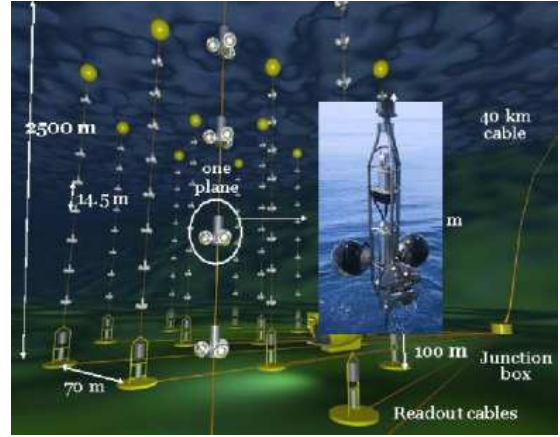}
\begin{quote} 
\caption{\small a) The underground Gran Sasso lab. Each of the 3 Halls, A, B, C, is $\sim$100m long. The CNGS $\nu$ beam arrives from the top of the figure. b) Layout of the Antares neutrino telescope in the Mediterranean sea, $\sim$40 km from Toulon.}
\label {fig:9}
\vspace{-0.6cm}
\end{quote}
\end{center}
\end{figure}

 It may be worth recalling that the MACRO \cite{25o} and SuperK experiments searched for $sub$-$dominant$ $oscillations$ due to a possible Lorentz invariant violation (in this case there would be mixing between flavour and velocity eigenstates). They were able only to place strong upper limits to possible Lorentz violation parameters. Also $neutrino$ $decay$ could be an explanation for neutrino disappearance: it was investigated, in particular for solar $\nu_{e}$: no decay was observed and limits were obtained \cite{25o}.

{\bf Neutrino telescopes.} Neutrinos from the sun, from the earth and from a supernova have been detected. Now one would like to detect high energy muon neutrinos coming from the same sources where cosmic rays are accelerated or from where come high energy gamma rays \cite{a}. Neutrinos interact rarely and may thus offer the possibility $to$ $see$ $the$ $sky$ in a different way; but one needs very large $neutrino$ $telescopes$. The Baikal neutrino telescope was one of the first to operate. A neutrino telescope is operating under the south pole ice (Amanda) and a km$^{3}$ size telescope is under construction there (ICECUBE); another neutrino telescope is operating in the Mediterranean sea (ANTARES Fig 11b) and a km$^{3}$ cube telescope is planned for the Mediterranean sea \cite{27n}. These telescopes are complementary: the south pole ice telescopes are looking at the northern sky, while the Mediterranean telescopes are looking at the southern sky, including the center of our galaxy.

 We discussed several times with Ben the outreach to be made from our experiments and from our studies. We dedicate to Ben the outreach lecture presented at a recent conference \cite{28n}. 

\section{Conclusions}

 In the past 50 years a great progress was made in particle physics and in astrophysics: an even greater progress was made in the understanding of the deep connections between the extremely small, the very large and the beginning of the universe. But we now know that the Standard Model of Particle Physics has limitations, that the universe contains a large fraction of dark matter and an even larger fraction of dark energy, all of which are yet to be understood.

 We must thank people like Ben who contributed to make the great progress of the past and who clearly were eager and had the knowledge to be ready for the tasks of the future.   

\section{Acknowledgments}

 We acknowledge several discussions with many colleagues, in particular at UCR, Bologna and CERN. We thank dr. M. Errico, R. Giacomelli, M. Giorgini and V. Togo for their cooperation.

\begin{figure}[!t]
\begin{center}
\includegraphics[width=.48\textwidth]{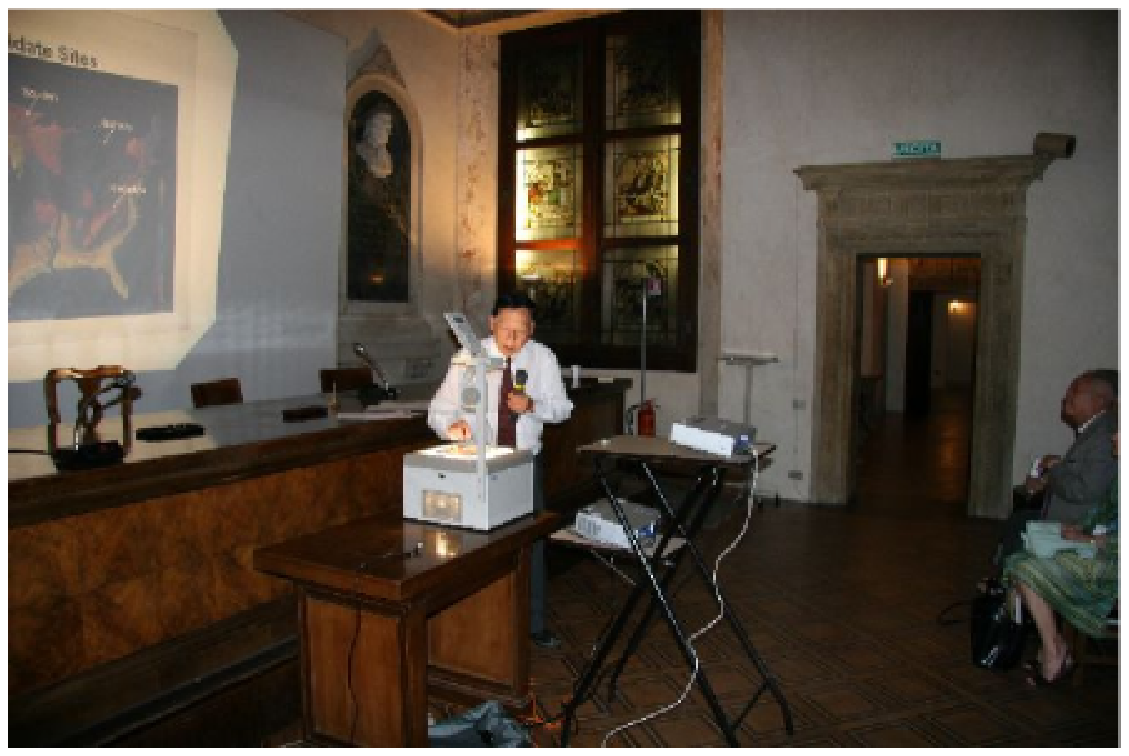}
\includegraphics[width=.48\textwidth]{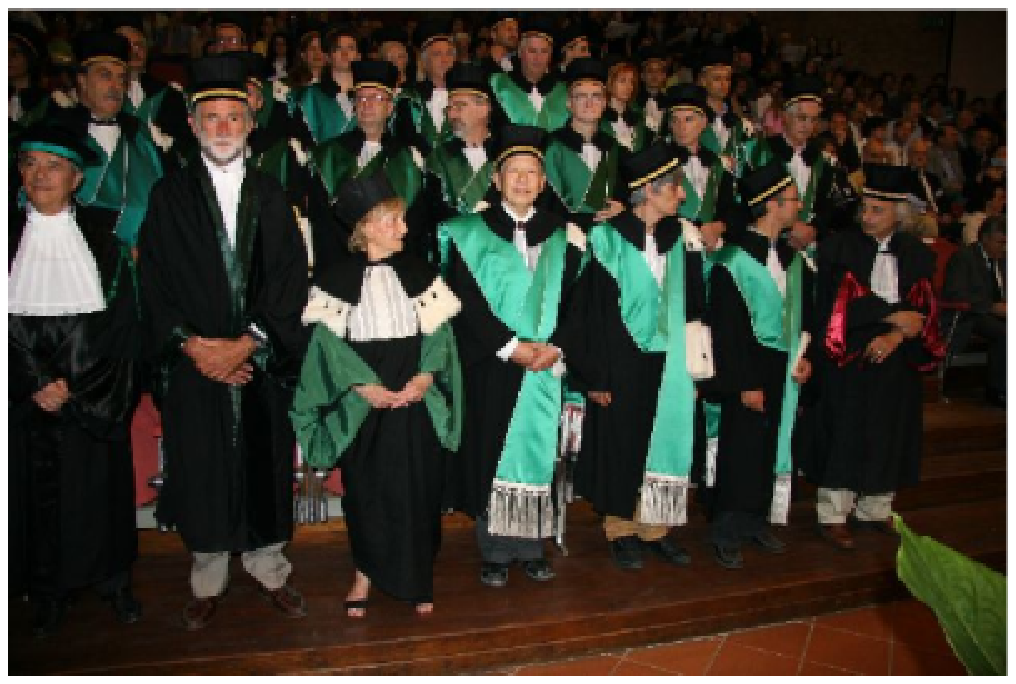}
\hspace{1cm}
\begin{quote} 
\caption{\small a) Ben Shen lecturing at the Bologna Academy of Sciences (he was a foreign corresponding member of the Academy). b) Ben at a ceremony at the University of Bologna.}
\label {fig:8a8b}
\vspace{-0.5cm}
\end{quote}
\end{center}
\end{figure}

\bibliographystyle{plain}

\end{document}